\documentclass[aps,prl,reprint,showpacs,groupedaddress,footinbib,citeautoscript]{revtex4-1}

\usepackage{graphicx}
\usepackage{amsmath}
\usepackage{amssymb}
\usepackage{mathrsfs}
\usepackage{bm}
\usepackage{array}
\newcolumntype {s}[1]{@{\hspace{#1}}} 
\usepackage[usenames,dvipsnames]{xcolor}
\usepackage{wrapfig}
\usepackage[normalem]{ulem}

\graphicspath{{.}{./FinalFigures/}{./EPS/}}

\newcommand* {\ee}{\mathrm{e}}

\newcommand*{\vek}[1]{{\bm{\mathrm{#1}}}}

\newcommand*{\kk}{{\bm{\mathrm{k}}}}

\newcount\colveccount
\newcommand*\colvec[1]{
        \global\colveccount#1
        \begin{pmatrix}
        \colvecnext
}
\def\colvecnext#1{
        #1
        \global\advance\colveccount-1
        \ifnum\colveccount>0
                \\
                \expandafter\colvecnext
        \else
                \end{pmatrix}
        \fi
}

\def\lsim{\raise0.3ex\hbox{$\;<$\kern-0.75em\raise-1.1ex\hbox{$\sim\;$}}}
\def\gsim{\raise0.3ex\hbox{$\;>$\kern-0.75em\raise-1.1ex\hbox{$\sim\;$}}}

\DeclareMathSymbol{\myRe}{\mathord}{symbols}{"3C}

\DeclareMathSymbol{\myIm}{\mathord}{symbols}{"3D}

\begin{document}

\title{Quantum capacitance of an HgTe quantum well as an indicator of the topological phase}

\author{T. Kernreiter}
\affiliation{School of Chemical and Physical Sciences and MacDiarmid Institute
for Advanced Materials and Nanotechnology, Victoria University of Wellington,
PO Box 600, Wellington 6140, New Zealand}

\author{M. Governale}
\affiliation{School of Chemical and Physical Sciences and MacDiarmid Institute
for Advanced Materials and Nanotechnology, Victoria University of Wellington,
PO Box 600, Wellington 6140, New Zealand}

\author{U. Z\"ulicke}
\affiliation{School of Chemical and Physical Sciences and MacDiarmid Institute
for Advanced Materials and Nanotechnology, Victoria University of Wellington,
PO Box 600, Wellington 6140, New Zealand}

\date{\today}

\begin{abstract}

Varying the quantum-well width in an HgTe/CdTe heterostructure allows to realize normal and
inverted semiconducting band structures, making it a prototypical system to study two-dimensional
(2D) topological-insulator behavior. We have calculated the zero-temperature thermodynamic density
of states $D_\mathrm{T}$ for the electron-doped situation in both regimes, treating interactions within
the Hartree-Fock approximation. A distinctively different behavior for the density dependence of
$D_\mathrm{T}$ is revealed in the inverted and normal cases, making it possible to detect the system's
topological phase through measurement of macroscopic observables such as the quantum
capacitance or electronic compressibility. Our results establish the 2D electron system in
HgTe quantum wells as unique in terms of its collective electronic properties.

\end{abstract}


\maketitle

\textit{Introduction.}---Capacitance measurements are a premier tool to elucidate the electronic
properties of two-dimensional (2D) electron systems~\cite{Smith1985,Luryi1988,Kravchenko1990,
Eisenstein1992PRL,Eisenstein1994PRB,Shapira1996,Millard1997PRB,Dultz2000,Allison2006,
MartinNP2008,Henriksen2010PRB,Li2011Sci,Young2012PRB,Kozlov2015PRL}. They fundamentally
probe the thermodynamic density of states,
\begin{equation}\label{eq:thdyDOS}
D_\mathrm{T} = \frac{\partial n}{\partial\mu}\quad ,
\end{equation}
where $n$ and $\mu$ denote the 2D system's electronic sheet density and chemical potential,
respectively. More specifically, $D_\mathrm{T}$ is related to the quantum capacitance per unit area
$C_\mathrm{q}$ and the electronic compressibility $K$ via
\begin{subequations}
\begin{eqnarray}
C_\mathrm{q} &=& e^2\, D_\mathrm{T} \quad , \\
K &=& \frac{D_\mathrm{T}}{n^2} \quad .
\end{eqnarray}
\end{subequations}
The intriguing interplay between single-particle and Coulomb-interaction contributions to
$D_\mathrm{T}$ has been intensely studied theoretically, both for conventional 2D electron systems
realized in heterostructures~\cite{vignalebook,Skinner2010a,Skinner2010b,Li2011PRB} and few-layer
graphene~\cite{Hwang2007PRL,Kusminskiy2008PRL,Borghi2010PRB,Abergel2011PRB,Li2011PRB}.
In particular, the tendency towards negative electronic compressibility in the low-density
limit~\cite{Tanatar1989PRB} has attracted a lot of attention~\cite{Kravchenko1990,Eisenstein1992PRL,
Eisenstein1994PRB,Millard1997PRB,Shapira1996,Dultz2000,Allison2006}.

Here we show how the thermodynamic density of states of electrons in an HgTe quantum well exhibits
behavior different from any of the previously studied 2D electron systems, essentially because of the
anomalous properties of an interaction-related inter-band contribution relevant for narrow-gap
systems. Our work provides new insight complementing the observation of unusual electric-transport 
properties in this system~\cite{Koenig2007Scien,Roth2009Scien,Brune2012NatPhys,Hart2015arXiv}
that relate to the existence of an unconventional, inverted, 2D electronic band structure when the
quantum-well width $d$ is larger than a critical value $d_{\text{c}}\approx 6.3\,$nm~\cite{Bernevig2006,
Koenig2008JPSJ,Winkler2012SSC,Tarasenko2015PRB}. The deeper understanding derived from our
results also enables novel characterization of topological phases~\cite{Qi2011RMP} in other
2D~\cite{Liu2008PRLa,Knez2011PRL} and bulk~\cite{Qi2011RMP,Hasan2011AnnuRev} materials and
extends the general knowledge about unusual collective properties of topological  and Dirac-semimetal
systems~\cite{Juergens2014PRL,Juergens2014PRB,Kernreiter2016PRX}.

\begin{figure}[b]
\includegraphics[width=7.5cm]{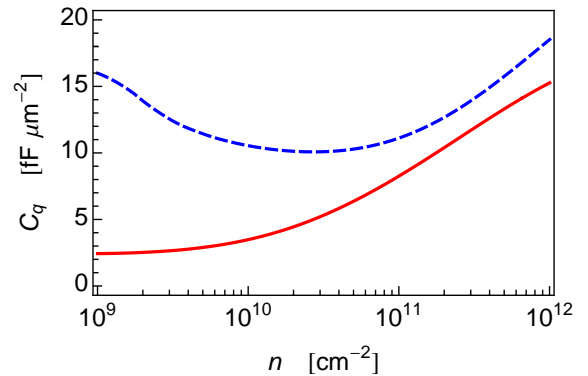}
\caption{\label{fig:QCapac}
Density dependence of the quantum capacitance per unit area for electrons in an HgTe 
quantum well. The red solid (blue dashed) curve is obtained for a quantum-well width $d=7\, $nm ($5\,
$nm) corresponding to the topological (normal) situation. Clearly distinguishable opposite trends
emerge in the low-density regime.}
\end{figure}

We calculate the thermodynamic density of states for electrons in HgTe quantum wells, taking
Coulomb interactions into account within the Hartree-Fock approximation. To be specific, we
focus on two experimentally feasible situations with quantum-well widths $d=5\,$nm and $7\,$nm, 
respectively, and present predictions for $D_\mathrm{T}$ as a function of the 2D-system's Fermi
wave vector. In our calculations, crucial effects arising from the finite width of electronic bound
states in the HgTe/CdTe heterostructure are included. Quite generally, we find that interaction
contributions significantly affect $D_\mathrm{T}$ and, thus, observables such as the quantum
capacitance and the electronic compressibility. See Fig.~\ref{fig:QCapac} for a pertinent example.
More specifically, it turns out that the inter-band exchange correction depends strongly on the
quantum-well width and changes its sign for a value close to $d_{\text{c}}$. We elucidate the
underlying mechanisms such as the interplay of band-structure parameters that lead to this
interesting behavior.

\textit{Model and Formalism.}---The theoretical framework for our calculation of many-particle effects
for electrons in an HgTe quantum well is based on the BHZ Hamiltonian~\cite{Bernevig2006}. The
latter adequately describes the relevant single-particle states in the low-energy band structure, using
basis functions $|E_1\pm\rangle$, which are superposition of conduction-electron and light-hole (LH)
states, and the heavy-hole (HH) states $|H_1\pm\rangle$. Within the representation defined by the
basis-state vector ($|E_1+\rangle, |H_1+\rangle, |E_1-\rangle, |H_1-\rangle$), the BHZ Hamiltonian
is block-diagonal and given by
\begin{eqnarray}\label{eq:BHZHam}
\mathscr{H}_0=
\begin{pmatrix} 
\mathcal{H}^{(+)} & 0  \\[2mm]
0&\mathcal{H}^{(-)} 
\end{pmatrix} \quad ,
\label{eq:BHZ}
\end{eqnarray}
with $\mathcal{H}^{(s)}=h^{(s)}_\mu\sigma^\mu$, $h^{(s)}=(C-Dk^2,s A k_x,-A k_y, M-B k^2)$ and
$\sigma^\mu=(\openone, \sigma_x, \sigma_y, \sigma_z)$ where $\sigma_j$ are the Pauli matrices.
The quantum number $s=\pm 1$ distinguishes spin-1/2 projections parallel to the quantum-well
growth direction, and the effective band-structure parameters $A, B, C, D, M$ are functions of the
quantum-well width $d$~\cite{Muehlbauer2014PRL}. For simplicity, we set the irrelevant overall
energy shift $C$ to zero. The sign of the gap parameter $M$ distinguishes the ordinary and
inverted-band situations: using the convention $B<0$, the system is in the topological (normal)
regime when $M<0$ ($M>0$).

The energy eigenvalues of the BHZ Hamiltonian (\ref{eq:BHZHam}) are given by~\cite{Bernevig2006}
\begin{equation}\label{eq:disper}
E_{\kk\alpha}^{(s)}\equiv E_{k\alpha}^{(s)}
= -D k^2 + \alpha\sqrt{(M-Bk^2)^2 + A^2 k^2} \quad ,
\end{equation}
where $\alpha=\pm 1$ distinguishes conduction and valence bands, both of which are doubly
degenerate in $s$. Due to the inherent axial symmetry of the BHZ model, the eigenvectors of the two
$2\times 2$ matrices $\mathcal{H}^{(s)}$ in Eq.~(\ref{eq:BHZHam}) can be expressed as
$a^{(s)}_{\kk\alpha}=U^{(s)}_{\phi_\kk} ~a^{(s)}_{k\alpha}$ in terms of the polar coordinates $(k,
\phi_\kk)$ for wave vector $\kk$, with
\begin{equation}\label{eq:BHZeigenvec}
a^{(s)}_{k\alpha} = \frac{1}{\sqrt{2}}
\begin{pmatrix}
\alpha \left[1 -\frac{ \alpha(Bk^2-M) }{\sqrt{A^2 k^2 + (Bk^2-M)^2}}\right]^{\frac{1}{2}}
\\[3mm]
s \left[1 +\frac{ \alpha(Bk^2-M) }{\sqrt{A^2 k^2 + (Bk^2-M)^2}}\right]^{\frac{1}{2}}
\end{pmatrix}
\end{equation}
and $U^{(s)}_{\phi_\kk}={\text{diag}}(\ee^{i s\phi_\kk/2},\ee^{-i s\phi_\kk/2})$.

\textit{Quantum many-body effects.}---The single-particle band dispersions given in 
Eq.~(\ref{eq:disper}) are renormalized by interaction effects. Assuming that the electrostatic
(Hartree) terms are compensated by the influence of a neutralizing background charge density, we
focus here on the exchange (Fock) contributions. The fundamental quasi-2D character of the charge 
carriers is accounted for by retaining the full $z$ dependence of quantum-well bound states through
the basis functions $|E_1+\rangle, |H_1+\rangle, |E_1-\rangle, |H_1-\rangle$ for the BHZ Hamiltonian.
The Fock self-energy of conduction-band electrons can then be written as 
\begin{widetext}
\begin{equation}\label{eq:Fockself}
\Sigma_{k\pm}^{(s)} = -2\pi C \int \frac{d^2k'}{(2\pi)^2} \,\, n_\mathrm{F}\big(E_{\kk'\pm}^{(s)}\big)
\int dz \int dz'  \,\, \frac{\ee^{-|\kk-\kk'||z-z'|}}{|\kk-\kk'|} \,\, \left[\psi^{(s)}_{\kk'\pm}(z)^\dagger\cdot
\psi^{(s)}_{\kk+}(z)\right] \left[\psi^{(s)}_{\kk+}(z')^\dagger\cdot\psi^{(s)}_{\kk'\pm}(z')\right] \quad ,
\end{equation}
\end{widetext}
where $C= e^2/(4\pi\epsilon\epsilon_0)$ measures the Coulomb-interaction strength, $n_\mathrm{F}
(E)$ is the Fermi function, and the  $\psi^{(s)}_{\kk\alpha}(z)$ are six-dimensional spinor wave
functions comprising the bands with $\Gamma_6$ and $\Gamma_8$ symmetry closest to the
bulk-material's fundamental gap~\cite{Winkler2003Book}. Intra-(inter-)band contributions to the Fock
self-energy are labeled by the subscript $+$ $(-)$. Note that terms with $s\neq s'$ vanish for the
block-diagonal BHZ model given above because of the orthogonality of the associated basis states.
However, such contributions do arise when spin-orbit-coupling effects are included. Effects of the
latter will be discussed briefly at the end of this paper.

In the zero-temperature limit, which we consider in the following, the Fermi functions in 
Eq.~(\ref{eq:Fockself}) reduce to $n_\mathrm{F}\big(E_{\kk-}^{(s)}\big)=1$ for the fully occupied
valence band and $n_\mathrm{F}\big(E_{\kk+}^{(s)}\big)=\Theta(k_\mathrm{F}-|\kk|)$, where
$k_\mathrm{F}$ is the modulus of the Fermi wave vector for electrons in the conduction band, and
$\Theta(\cdot)$ denoted the Heaviside step function. To take into account both the in-plane dynamics
described by the BHZ Hamiltonian as well as the nontrivial spinor structure of the BHZ-model basis
states, we employ subband $\kk\cdot\vek{p}$ theory~\cite{bro85,bro85a} to write the spinor wave
functions $\psi^{(s)}_{\kk\alpha}(z)$ as superpositions
\begin{equation}\label{eq:kdepspinors}
\psi^{(s)}_{\kk\alpha}(z)=
\sum^2_{i=1}  
\left(U^{(s)}_{\phi_\kk}\right)_{ii}a_{k\alpha, i}^{(s)}~\psi_{0i}^{(s)}(z) \quad ,
\end{equation} 
where the coefficients $a_{k\alpha, i}^{(s)}$ are the components of the corresponding eigenvectors, 
Eq.~(\ref{eq:BHZeigenvec}), of the BHZ Hamiltonian. The six-dimensional spinors $\psi^{(s)}_{0i}(z)$ 
are the BHZ-model basis-state spinors for zero in-plane wave vector, which are determined by the 
solutions to a confined-particle problem for the HgTe/CdTe quantum-well heterostructure. Their explicit
expressions have been given in the supplemental information of Ref.~\cite{Bernevig2006}, where for
instance $\psi^{(+)}_{01}(z)^T=(f_1(z),0,0,f_4(z),0,0)$ and $\psi^{(+)}_{02}(z)^T=(0,0,f_3(z),0,0,0)$
which are normalized, i.e., $\int dz~|\psi^{(s)}_{0i}(z)|^2=1$, $\forall~i,s$. As a result, we obtain for
the intra- and inter-band contributions to the Fock self-energy
\begin{eqnarray}\label{eq:Fockself2}
\Sigma_{k\pm}^{(s)}&=&\frac{-C}{\pi}\int_0^{\pi}\! d\phi\int_0^{k_\pm} \!\! dk' \, k'\int \! dz \int \! dz' 
\,\, \frac{\ee^{-r(k,k',\phi)|z-z'|}}{r(k,k',\phi)}\nonumber\\[2mm]
&&{}\hskip-0.9cm\times\sum_{i,j}\mathcal{F}_{ij}(\phi)~a_{k'\pm,i}^{(s)}a_{k'\pm,j}^{(s)}a_{k+,i}^{(s)}
a_{k+,j}^{(s)}|\psi_{0i}^{(s)}(z)|^2|\psi_{0j}^{(s)}(z')|^2,\nonumber\\
\end{eqnarray}
where the integration limits are $k_+=k_\mathrm{F}$ (intra-band) and $k_-=k_{\text{c}}$ (inter-band),
with $k_{\text{c}}$ being an ultraviolet cutoff. In Eq.~(\ref{eq:Fockself2}), $r(k,k',\phi)=\sqrt{k^2+k'^2-
2k\, k'\cos\phi}$ and $\mathcal{F}_{ij}(\phi)=\sqrt{1-(1-\delta_{ij})\sin^2\phi}$, with $\phi\equiv \phi_\kk
-\phi_{\kk'}$ and $\delta_{ij}$ being the Kronecker symbol. The inter-band contribution depends
logarithmically on $k_{\text{c}}$, which is typically chosen to be of the order of the inverse lattice
constant~\cite{Hwang2007PRL,Kusminskiy2008PRL}. Finally, with the chemical potential given in
terms of $k_\mathrm{F}$ as $\mu=E_{k_\mathrm{F}+}^{(s)}+\Sigma_{k_\mathrm{F}+}^{(s)}+
\Sigma_{k_\mathrm{F}-}^{(s)}$, and using the relation $n=k_\mathrm{F}^2/(2\pi)$, the expression
(\ref{eq:thdyDOS}) for the thermodynamic density of states can be rewritten as
$D_\mathrm{T} = \left( \frac{\pi}{k_\mathrm{F}}\, \frac{\partial \mu}{\partial k_\mathrm{F}} \right)^{-1}$.
Measuring wave vectors and energies in terms of the BHZ-model scales $q_0\equiv A/|B|$ and
$E_0\equiv Aq_0$, the natural unit for $D_\mathrm{T}^{-1}$ is $|B|$. The fine-structure constant that
appears in the exchange-energy contributions to $\mu$ is given by $\alpha_{\text{qw}}\equiv e^2/(4\pi
\epsilon\epsilon_0  A) \approx 0.19$ when using $\epsilon=20.8$ as the dielectric constant of HgTe.

\begin{table}[b]
\begin{tabular}{|l||c|c|}
\hline
& $d=5\,$nm & $d=7\,$nm   \\ \hline\hline
$A\,\, [\mathrm{eV\, nm}]$ & 0.365 & 0.340 \\ \hline
$B\,\, [\mathrm{eV\, nm}{}^2]$ & -0.50 & -0.50 \\ \hline
$D\,\, [\mathrm{eV\, nm}{}^2]$ & -0.50 & -0.87 \\ \hline
$M\,\, [\mathrm{meV}]$ & 24.0 & -8.5  \\ \hline
\end{tabular}
\caption{\label{tab:Input} 
Parameters of the BHZ model applicable for two experimental realizations of HgTe quantum
wells~\cite{Muehlbauer2014PRL} having widths $d=5\,$nm and $7\,$nm, respectively.}
\end{table}

\begin{figure}[t]
\includegraphics[width=7.5cm]{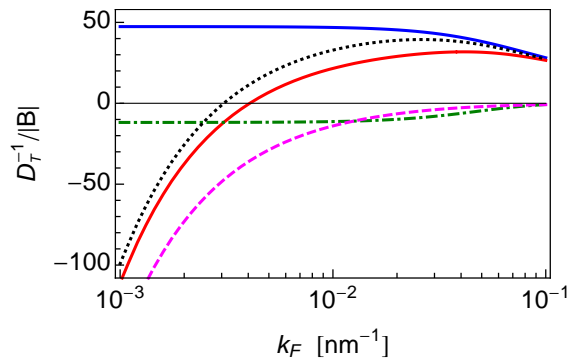}
\caption{\label{fig:Compressnor}
Inverse thermodynamic density of states $D_\mathrm{T}^{-1}\equiv \partial\mu/\partial n$ of an HgTe
quantum well in the normal regime (well width $d=5\,$nm). The red (blue) solid curve shows
the result with (without) interactions. The magenta dashed (green dot-dashed) curve is the intra-band
(inter-band) exchange contribution only. The black dotted curve is the sum of the noninteracting
and intra-band exchange contributions.}
\end{figure}

\textit{Numerical results for $D_\mathrm{T}$.}---We now present results obtained for the
thermodynamic density of states in normal and topological HgTe quantum wells. Following the
usual convention, $D_\mathrm{T}^{-1}\equiv \partial\mu/\partial n$ is shown
as a function of the Fermi wave vector. We first consider an HgTe quantum well with width
$d=5\,$nm, which is in the the normal (non-inverted band-structure) regime. The associated BHZ
parameters are given in Table~\ref{tab:Input} and correspond to an actual experimental
realization~\cite{Muehlbauer2014PRL}. For the large-momentum cutoff of the inter-band contribution,
we choose $k_{\text{c}}=a_0^{-1}$, with $a_0=0.646\,$nm being the HgTe bulk-material lattice constant.
We show the result obtained for $D_\mathrm{T}^{-1}$ in Fig.~\ref{fig:Compressnor}, making also explicit
the various contributions to $D_\mathrm{T}^{-1}$.  The purely kinetic (i.e.,
noninteracting) part is given by a constant in the low-density regime,
\begin{equation}\label{eq:kineticpart}
\left.\frac{\partial E_{k_\mathrm{F}\alpha}^{(s)}}{\partial n}\right|_{k_\mathrm{F}=0}=2\pi\left[
\alpha\left(\frac{A^2}{2|M|}+|B|\, \text{sign}(M)\right) - D \right] ,
\end{equation}
which has the form expected for an ordinary 2D electron system~\footnote{ The r.h.s.\ of
Eq.~(\ref{eq:kineticpart}) can be expressed as $\pi\hbar^2/m_\ast$, where $m_\ast$ is the effective
band mass of 2D electrons obtained from the small-$k$ expansion of the dispersion (\ref{eq:disper}).}.
However, it exhibits a weak dependence on $k_\mathrm{F}$  at larger carrier densities due to the HH-LH
mixing of quantum-well bound states having finite in-plane wave vector. The intra-band interaction (Fock)
renormalization term is always negative and therefore reduces $D_\mathrm{T}^{-1}$, thus leading to an
enhancement of the electronic compressibility. At low-enough densities, the intra-band contribution drives
$D_\mathrm{T}^{-1}$ to negative values. Such a behavior is also reminiscent of that of an ordinary 2D
electron system~\footnote{The intra-band Fock contribution to $D_\mathrm{T}^{-1}$ diverges $\propto
k_\mathrm{F}^{-1}$ in the low-density limit like the Fock contribution for an ordinary 2D electron
gas~\cite{Skinner2010b}. In real samples, this divergence is cut off by image-charge
effects~\cite{Skinner2010a,Skinner2010b}.}. In the normal regime (except very close to the critical well
width $d_c$), the inter-band exchange contribution is also negative and thus reduces
$D_\mathrm{T}^{-1}$ further.  As a result, the crossover from positive to negative values of
$D_\mathrm{T}^{-1}$ is shifted to higher densities. This behaviour has to be contrasted to that exhibited
by single-layer graphene where the exchange renormalization of $D_\mathrm{T}^{-1}$ is
positive~\cite{Hwang2007PRL,Kusminskiy2008PRL}. Overall, from the results shown in
Fig.~\ref{fig:Compressnor}, we see that the exchange contributions strongly  influence the electronic
compressibility. 

\begin{figure}[t]
\includegraphics[width=7.5cm]{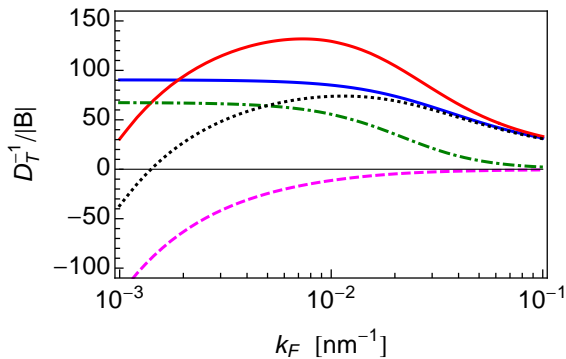}
\caption{\label{fig:Compresstop}
Inverse thermodynamic density of states $D_\mathrm{T}^{-1}\equiv \partial\mu/\partial n$ of an HgTe
quantum well in the inverted regime (well width $d=7\,$nm). The red (blue) solid curve shows
the result with (without) interactions. The magenta dashed (green dot-dashed) curve is the intra-band
(inter-band) exchange contribution only. The black dotted curve is the sum of the noninteracting
and intra-band exchange contributions. Notice the opposite sign of the inter-band exchange contribution
(green dot-dashed curve), which shifts the crossover to negative compressibility to very
low carrier densities.}
\end{figure}

We now consider the inverted regime of an HgTe quantum well, which is realized for a well width
$d>d_{\text{c}}\approx 6.3\,$nm. Taking the BHZ parameters of a feasible experimental situation
corresponding to a well width $d=7\,$nm (see Table~\ref{tab:Input}), we again calculate 
the quantity $D_\mathrm{T}^{-1}$. The result is shown in Fig.~\ref{fig:Compresstop}. The most salient
feature is that the inter-band exchange contribution is now \emph{positive\/}, like in single-layer
graphene~\cite{Hwang2007PRL,Kusminskiy2008PRL}, and considerably larger in magnitude
as compared to the situation in the normal regime. In contrast, the intra-band exchange term
is of similar magnitude and has the same sign as in the normal case. The kinetic (noninteracting)
contribution is much larger as compared to the $d=5$-nm case, which is mainly due to the smaller
band gap in the present case --- this can be inferred from Eq.~(\ref{eq:kineticpart}). We see that
the electronic compressibility is reduced by up to 35\% due to exchange effects as compared with
the noninteracting case. This trend is changed only at very low densities where the (negative)
intra-band contribution becomes dominant.  

\begin{figure}[t]
\vspace*{-0.4cm}
\includegraphics[width=7.5cm]{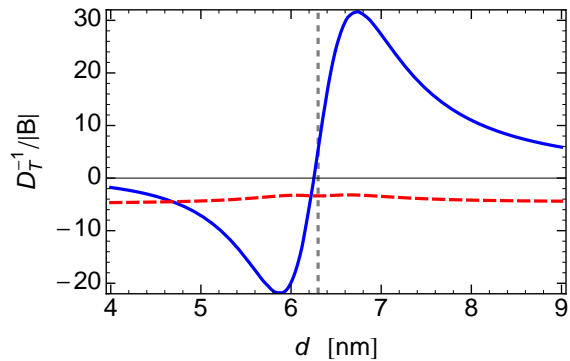}
\caption{\label{fig:CompressIntraInter}
Intra-band (dashed red curve) and inter-band (solid blue curve) exchange
contribution to the inverse thermodynamic density of states $D_\mathrm{T}^{-1}$ as a function of the
HgTe quantum-well width $d$ for carrier sheet density $n=10^{10\,}$cm${}^{-2}$. The dashed
vertical line indicates the value of the critical well width $d_{\text{c}}=6.3\,$nm.}
\end{figure}

The striking difference observed between the inter-band interaction-renormalization contributions
in the topological and normal regimes invites more detailed scrutiny. Figure~\ref{fig:CompressIntraInter}
illustrates the variation of intra-band and inter-band exchange terms as a function of the quantum-well
width \footnote{The BHZ parameters $A$ and $B$ are generally only weakly dependent on the HgTe
quantum-well width $d$ (see Table~\ref{tab:Input}) and, for this calculation, we use the $A$ and $B$
values for $d=7\,$nm. The mass parameter $M$, on the other hand, depends sensitively on $d$, and
we extract its functional dependence from the band-edge energies of the conduction and valence bands,
which are obtained by appropriate matching conditions derived from the relevant confined-electron
problem (see the supplemental information of Ref.~\onlinecite{Bernevig2006}). Note that the BHZ
parameter $D$ does not enter the exchange corrections to $D_\mathrm{T}^{-1}$.} for a fixed carrier
density $n=10^{10}\,$cm${}^{-2}$. The intra-band contribution is always negative and rather insensitive
to a variation of $d$. The inter-band contribution, however, depends strongly on the quantum-well width
and changes its sign in the vicinity of the critical value $d_{\text{c}}\approx 6.3\,$nm. Also around
$d_{\text{c}}$, due to the vanishing band gap, we can anticipate the onset of a divergence in the
inter-band contribution for $k_\mathrm{F}\to 0$. Figure~S1 in the Supplemental Material
shows this even more clearly.  We can attribute the sign change in the inter-band exchange contribution
to $D_\mathrm{T}^{-1}$ to a complex interplay of band-mixing effects (due to the terms proportional to
$A$ in the BHZ Hamiltonian) and the change of the band characters when crossing over from $M>0$ to
$M<0$. To be more specific, we find that the heavy-hole term ($i=j=2$) in Eq.~(\ref{eq:Fockself2}) gives
generally (especially for low densities) the largest contribution to $\Sigma_{k_\mathrm{F}-}^{(s)}$ (as well
as to $\partial\Sigma_{k_\mathrm{F}-}^{(s)}/\partial n$), where for $M>0$ ($M<0$) it is a monotonically
decreasing (increasing) function of $k_\mathrm{F}$.

\textit{Effect of spin-orbit coupling.}---We have extended our calculation of the thermodynamic density of
states to the situation with bulk-inversion-asymmetry and structural-inversion-asymmetry spin-orbit
coupling~\cite{Rothe2010NJP,Koenig2008JPSJ}  and find that, only for the largest expected magnitudes
of the bulk-inversion-asymmetry energy scale of a few meV, results change quantitatively by upto 10\%.
However, our findings suggest that spin-orbit coupling affects the electronic compressibility of electrons in
HgTe quantum wells typically only at the percent level. See the Supplemental Material for more details.

\textit{Conclusions.}---We have presented results for the thermodynamic density of states for
electrons in HgTe quantum wells in experimentally feasible situations. Interaction effects have
been included within the Hartree-Fock approximation. We have also taken into account the finite
width of the HgTe/CdTe quantum-well heterostructure, which is necessary to account for
the attenuated Coulomb repulsion in the transverse direction. Markedly different behavior is exhibited
for a well width of $d=5\,$nm (normal regime) compared to one with $d=7\,$nm (topological regime).
We have pinpointed the origin of this finding as the sizeable inter-band exchange correction whose
sign differs in the topological and normal regimes. Thus a measurement of the quantum capacitance
of HgTe quantum wells, e.g.\ using HgTe double-quantum-well configurations~\cite{Yakunin2016PRB},
provides a useful way to determine the topological state of this system.

The enhancement and eventual sign change of the compressibility found in the low-density limit of
the non-topological phase is analogous to the behavior exhibited by ordinary 2D electron systems with
parabolic dispersion~\cite{Kravchenko1990,Eisenstein1992PRL,Eisenstein1994PRB,Millard1997PRB,
Shapira1996,Dultz2000,Allison2006}. In contrast, the compressibility of the 2D electron system in the
topological phase is strongly suppressed by Coulomb interactions. Additional contributions to the
compressibility arising from image charges~\cite{Skinner2010a,Skinner2010b} and
disorder~\cite{Fogler2004PRB} can be straightforwardly included to facilitate the description of real
samples.

%
%


\setcounter{equation}{0}
\setcounter{figure}{0}
\setcounter{section}{0}
\makeatletter
\renewcommand{\theequation}{S\arabic{equation}}
\renewcommand{\thefigure}{S\arabic{figure}}
\renewcommand{\thesection}{S.\arabic{section}}

\newpage

\pagestyle{empty}

\section*{Supplemental Material for "Quantum capacitance of an HgTe quantum well as an indicator
of the topological phase"}

\section{Inter-band self-energy contribution}

\begin{figure}[h]
\includegraphics[width=6cm]{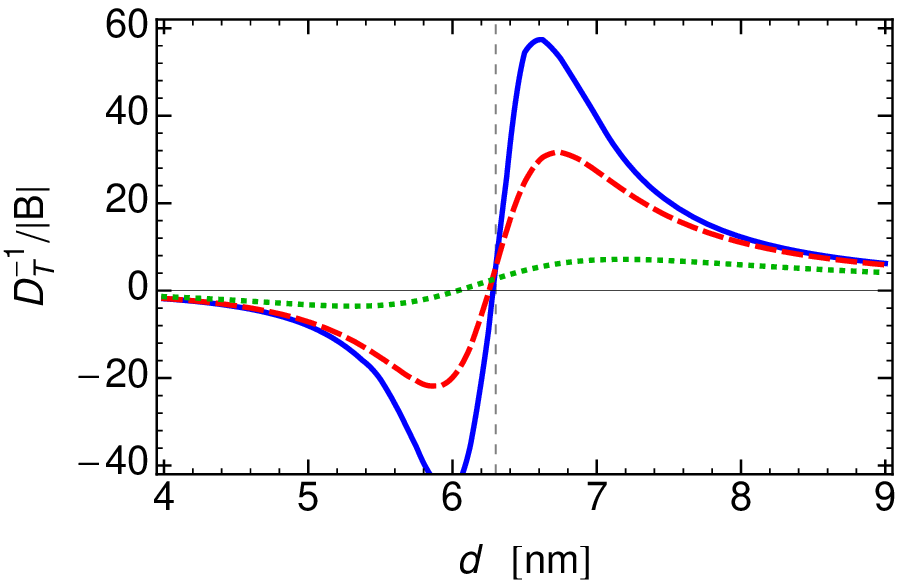}
\caption{\label{fig:Interwrtd}
Inter-band  exchange contribution to the inverse thermodynamic density of states $D_\mathrm{T}^{-1}$
as a function of the HgTe quantum-well width $d$ for carrier sheet densities $n=5\times 10^{9}\,$cm${}^{-2}$
(blue solid curve), $10^{10}\,$cm${}^{-2}$ (red dashed curve), and $5\times 10^{10}\,$cm${}^{-2}$ (green
dotted curve). The dashed vertical line indicates the value of the critical well width $d_{\text{c}}=6.3\,$nm.}
\end{figure}

We show in Fig.~\ref{fig:Interwrtd} the inter-band Fock
contribution to $D_\mathrm{T}^{-1}$ for three different densities $n=(5\times 10^{9},10^{10},5\times
10^{10})\,$cm${}^{-2}$. Especially for lower densities, it can be clearly seen that the inter-band
contribution changes its sign close to the critical value $d_{\text{c}}\approx 6.3\,$nm. For higher
densities, however, the crossover occurs at somewhat smaller values of $d$. The reason for this
tendency can be traced to the fact that, for larger densities, other terms besides the dominant
HH-related contributions become important in the sum in Eq.~(\ref{eq:Fockself2}). 

\section{Effects of spin-orbit coupling}

Spin-orbit coupling due to structural inversion asymmetry (SIA) and bulk inversion asymmetry
(BIA) can be straightforwardly incorporated into the BHZ Hamiltonian. In the following, we
discuss the influence of those types of spin-orbit coupling separately.  

\begin{figure}[t]
\includegraphics[width=6cm]{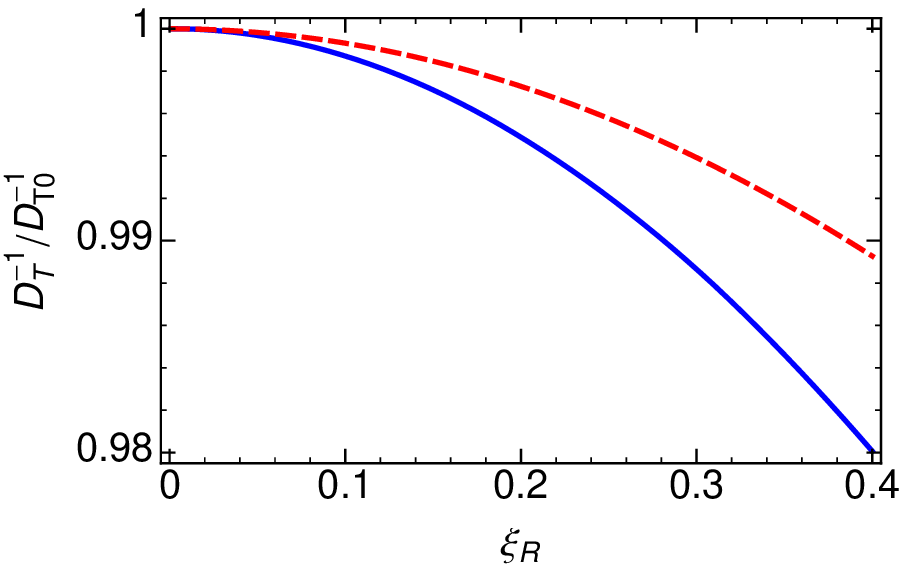}
\caption{\label{fig:SIA}
Dependence of $D_\mathrm{T}^{-1}/D_{\mathrm{T}0}^{-1}$ on the SIA magnitude measured in terms
of $\xi_{\text{R}}$. Here $D_\mathrm{T}$ ($D_{\mathrm{T}0}^{-1}$) is the thermodynamic density
of states obtained with (without) SIA for quantum-well width $d=7\,$nm and $n=5\times 10^{9}\,\
\mathrm{cm}^{-2}$ (red dashed curve) and $5\times 10^{10}\,\mathrm{cm}^{-2}$ (blue solid curve).}
\end{figure}

\textit{SIA\/}.
The leading contribution due to SIA arising from the presence of a perpendicular electric field
$\mathcal{E}_z$ is linear in the wave vector and given by~\cite{Rothe2010NJP}
\begin{equation}\label{eq:effSIA}
\mathscr{H}_{\text{R}}=
\begin{pmatrix} 
0 & 0 & -i R_0k_- & 0  \\[2mm]
0& 0& 0 & 0  \\[2mm]
i R_0k_+ & 0 & 0 & 0 \\[2mm]
0 & 0 & 0& 0
\end{pmatrix}~.
\end{equation}
Typical values are $R_0/(e\mathcal{E}_z)=-15.6\,$nm${}^2$~\cite{Rothe2010NJP} and
$\mathcal{E}_z=0.01\,\mathrm{V\, nm}{}^{-1}$~\cite{Hart2015arXiv}. We define the SIA-related
dimensionless parameter $\xi_{\text{R}} \equiv |R_0/A|$ ($\approx 0.44$ in a typical HgTe quantum
well~\cite{Rothe2010NJP}). Figure~\ref{fig:SIA} illustrates the effect of SIA on $D_\mathrm{T}^{-1}$
in the inverted regime ($d=7\,$nm), demonstrating that SIA reduces $D_\mathrm{T}^{-1}$ only slightly 
and that the reduction is larger for higher densities. However, the relative change amounts to a few
percent only. For the normal case ($d=5\,$nm), the change is even less than 1\%.

\begin{figure}[b]
\includegraphics[width=6cm]{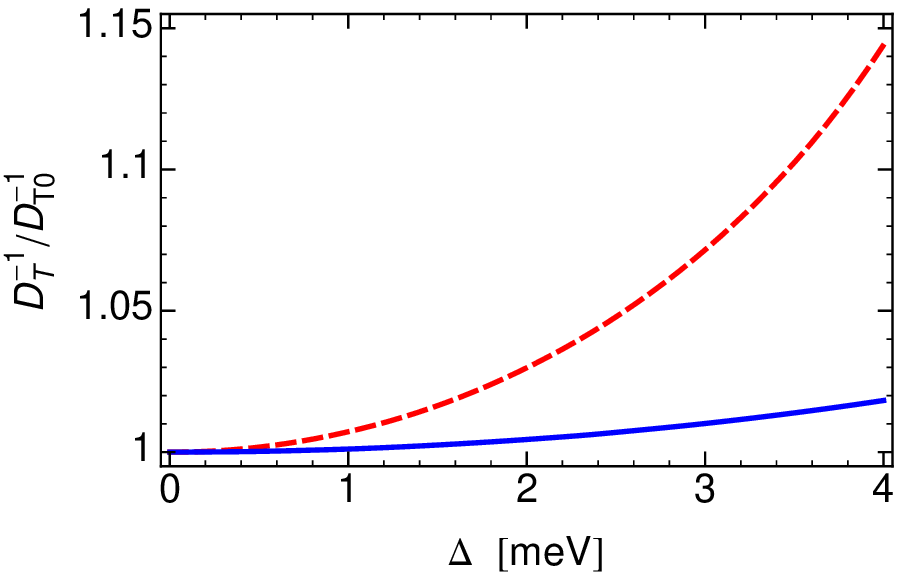}
\caption{\label{fig:BIA}
Dependence of $D_\mathrm{T}^{-1}/D_{\mathrm{T}0}^{-1}$ on the BIA magnitude $\Delta$. Here
$D_\mathrm{T}$ ($D_{\mathrm{T}0}^{-1}$) is the thermodynamic density of states obtained with
(without) BIA for quantum-well width $d=7\,$nm and $n=5\times 10^{9}\,\mathrm{cm}^{-2}$ (red
dashed curve) and $5\times 10^{10}\,\mathrm{cm}^{-2}$ (blue solid curve). In actual
samples, $\Delta=1.8\,$meV~\cite{Koenig2008JPSJ,Hart2015arXiv}.}
\end{figure}

\textit{BIA\/}.
The effect of BIA can be accounted for by augmenting the BHZ Hamiltonian, Eq.~(\ref{eq:BHZHam}),
by the term~\cite{Koenig2008JPSJ}
\begin{equation}\label{eq:effBIA}
\mathscr{H}_{\Delta}=
\begin{pmatrix} 
0 & 0 & 0 & -\Delta  \\[2mm]
0& 0& \Delta & 0  \\[2mm]
0 & \Delta & 0 & 0 \\[2mm]
- \Delta & 0 & 0& 0
\end{pmatrix}~.
\end{equation}
In Fig.~\ref{fig:BIA}, we plot the relative change of $D_\mathrm{T}^{-1}$ due to BIA in the inverted
regime. The figure shows that BIA leads to an increase of $D_T^{-1}$ by up to 15\%, which further
increases the difference between the magnitudes obtained for this quantity in the interacting and
noninteracting cases. Also, in contrast to SIA, the effect of BIA is larger for smaller densities.

\end{document}